\documentclass{elsart1p}
\usepackage{graphicx}
\usepackage{amssymb}
\newcommand{\beq}{\begin{equation}}
\newcommand{\eeq}[1]{\label{#1} \end{equation}} 
\begin{document}
\begin{frontmatter}
%
%
%
\title{Two and three nucleon $K^-$ absorption in nuclei}
%
%
\author[magas]{V.K.~Magas}, 
\author[oset]{E.~Oset},
\author[magas]{A.~Ramos}
\address[magas]{Departament d'Estructura i Constituents de la Mat\'eria,\\
              Universitat de Barcelona,  Diagonal 647, 08028 Barcelona, Spain} 
\address[oset]{Departamento de F\'{\i}sica Te\'orica and 
              IFIC Centro Mixto Universidad de Valencia-CSIC\\
              Institutos de Investigaci\'on de Paterna,
               Apdo. correos 22085, 46071, Valencia, Spain}

\begin{abstract}
We analyze the peaks in the ($\Lambda p$) and ($\Lambda d$) invariant mass 
distributions, observed in recent FINUDA experiments and claimed to be signals 
of deeply bound kaonic states, and find them to be naturally explained in 
terms of $K^-$ absorption by two or three nucleons leaving the rest of the 
target nucleus as a spectator. For reactions on heavy nuclei, the subsequent 
interactions of the particles produced in the primary absorption process 
with the residual nucleus play an important role. Thus at present there is 
no experimental evidence of deeply bound $K^-$ states in nuclei. 
\end{abstract}
\begin{keyword}
$K^-$ absorption in nuclei \sep many body absorption \sep final state 
interaction
%
\PACS 13.75.-n  \sep 13.75.Jz \sep 21.65.+f \sep 25.80.Nv
\end{keyword}
\end{frontmatter}
%
\section{$K^-$-nucleons bound states at FINUDA}
\label{sec-1}

The possibility of having deeply bound $K^-$ states in nuclei 
is receiving much attention both theoretically (see for example 
Ref.~\cite{Ramos:2008npa} for an overview) and experimentally.
Here we discuss recent FINUDA data that have been sometimes used 
to imply the existence of deeply bound $K^-$ states. 

\begin{figure}
\includegraphics[width=0.75\textwidth]{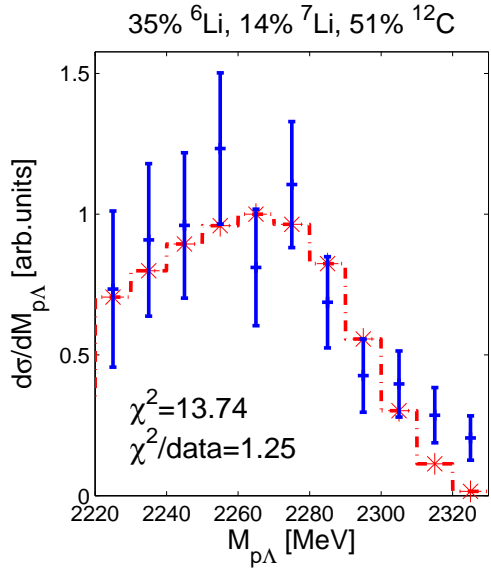} 
\caption{Invariant mass of $\Lambda p$ distribution for $K^-$ absorption 
in light nuclei in the following proportion \cite{Agnello:2005qj}:  
51\% $^{12}$C, 35\% $^{6}$Li and 14\% $^{7}$Li, including
kaon boost machine corrections \cite{future_finuda}. Stars and histogram 
show result of our calculations \cite{Magas:2006fn}, experimental points 
and error bars are taken from \cite{Agnello:2005qj}.}
\label{fig1}      
\end{figure} 

A peak observed by the FINUDA collaboration 
\cite{Agnello:2005qj} in the invariant mass distribution of $\Lambda p$, 
following $K^-$ absorption in a mixture of light nuclei, was interpreted 
as evidence for a $K^- pp$ bound state, with 115 MeV binding and 67 MeV 
width. However, it was soon shown in \cite{Magas:2006fn} that the peak 
seen is naturally explained in terms of $K^-$ absorption on a pair of 
nucleons leading to a $\Lambda p$ pair, followed by 
final state interactions (FSI), i.e. by the rescattering of 
$p$ or $\Lambda$ on the remnant nucleus, and that the back-to-back 
correlation of the $\Lambda p$ pairs is preserved to large extent. 
Fig.~\ref{fig1} shows the $\Lambda p$ invariant mass distribution for 
$K^-$ absorption in a mixture of light nuclei following the proportion 
\cite{Agnello:2005qj}: 51\% $^{12}$C, 35\% $^{6}$Li and 14\% $^{7}$Li, 
including kaon boost machine corrections \cite{future_finuda}. 
However, we stress that in this mixture the contribution of reactions with 
$^{12}$C is absolutely dominant, about 99\%. This is mostly due to the 
two orders of magnitude higher overlap of the nuclear density with the 
corresponding $K^-$ wavefunction $\int d\vec{r}\, |\Psi_{K^-}(\vec{r}\,)|^2 
\rho_{A}^2(r)$ in $^{12}$C than in Li. Nevertheless, in order to verify in 
detail our model of $K^-$ two nucleon absorption dynamics, our choice of 
$\Psi_{K^-}(\vec{r}\,)$, and our simulation of FSI, experimental spectra 
for separate nuclei are necessary, while at the moment the ($\Lambda p$) 
invariant mass distribution is only available for the mixture of the three 
lightest targets \cite{Agnello:2005qj}. 
        
More recently, a new experiment of the FINUDA 
collaboration \cite{:2007ph} found a peak on the invariant mass of 
$\Lambda d$ following the absorption of a $K^-$ on $^6$Li, which was 
interpreted as a signature for a bound $\bar{K}NNN$ state with 58 MeV 
binding and 37 MeV width. Note that this result is in disagreement with 
the previous FINUDA statement \cite{Agnello:2005qj}, since the bound state 
of the $K^-$ with three nucleons should have a larger binding energy than 
with two nucleons. 
    
A similar experiment was performed at KEK \cite{Suzuki:2007kn} on $^4$He 
target, looking at the $\Lambda d $ invariant mass following $K ^-$ 
absorption. It is claimed, however, that the observed peak could be 
a signature of three body absorption.  

\begin{figure*} 
\includegraphics[width=0.48\textwidth]{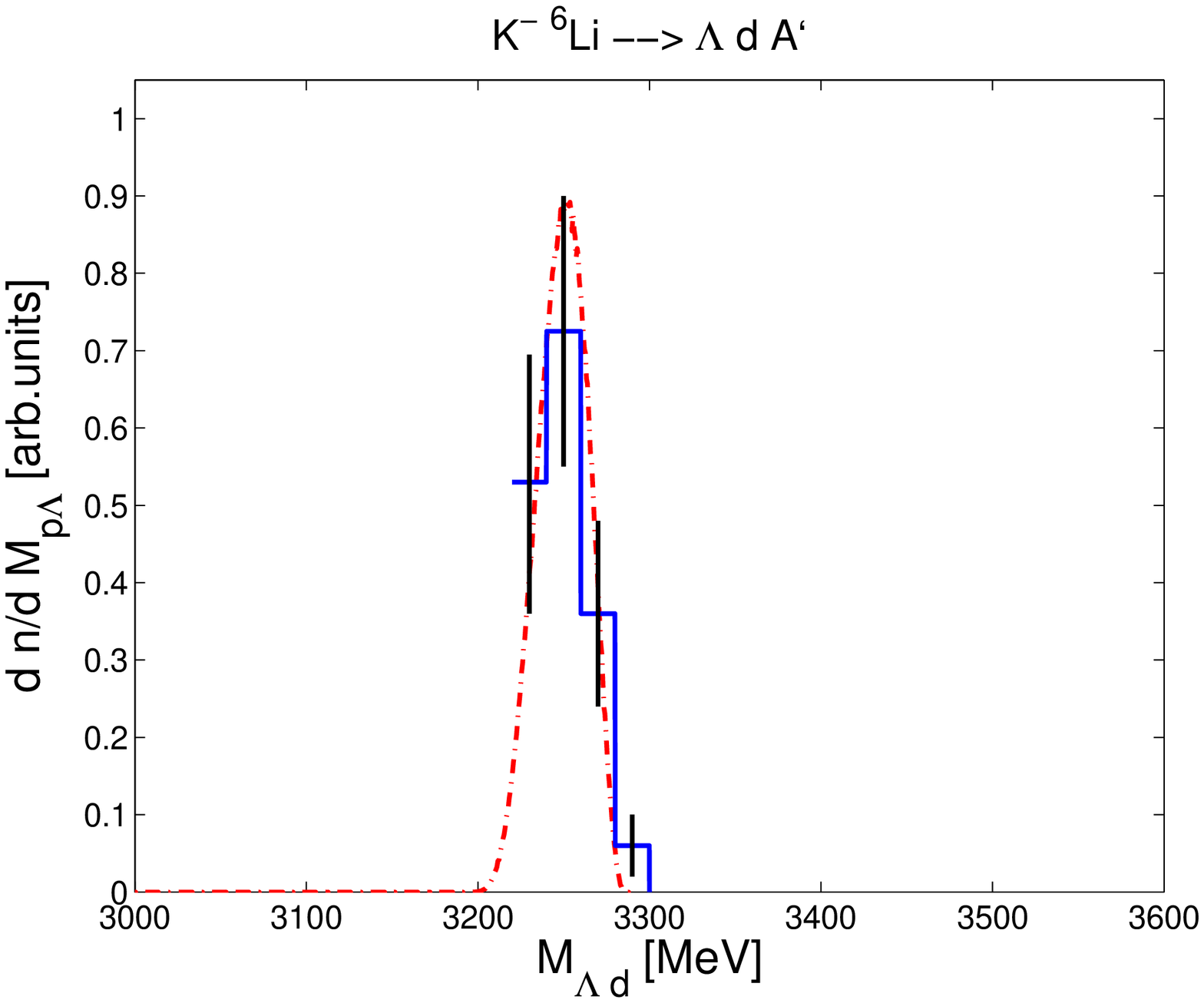} 
\includegraphics[width=0.48\textwidth]{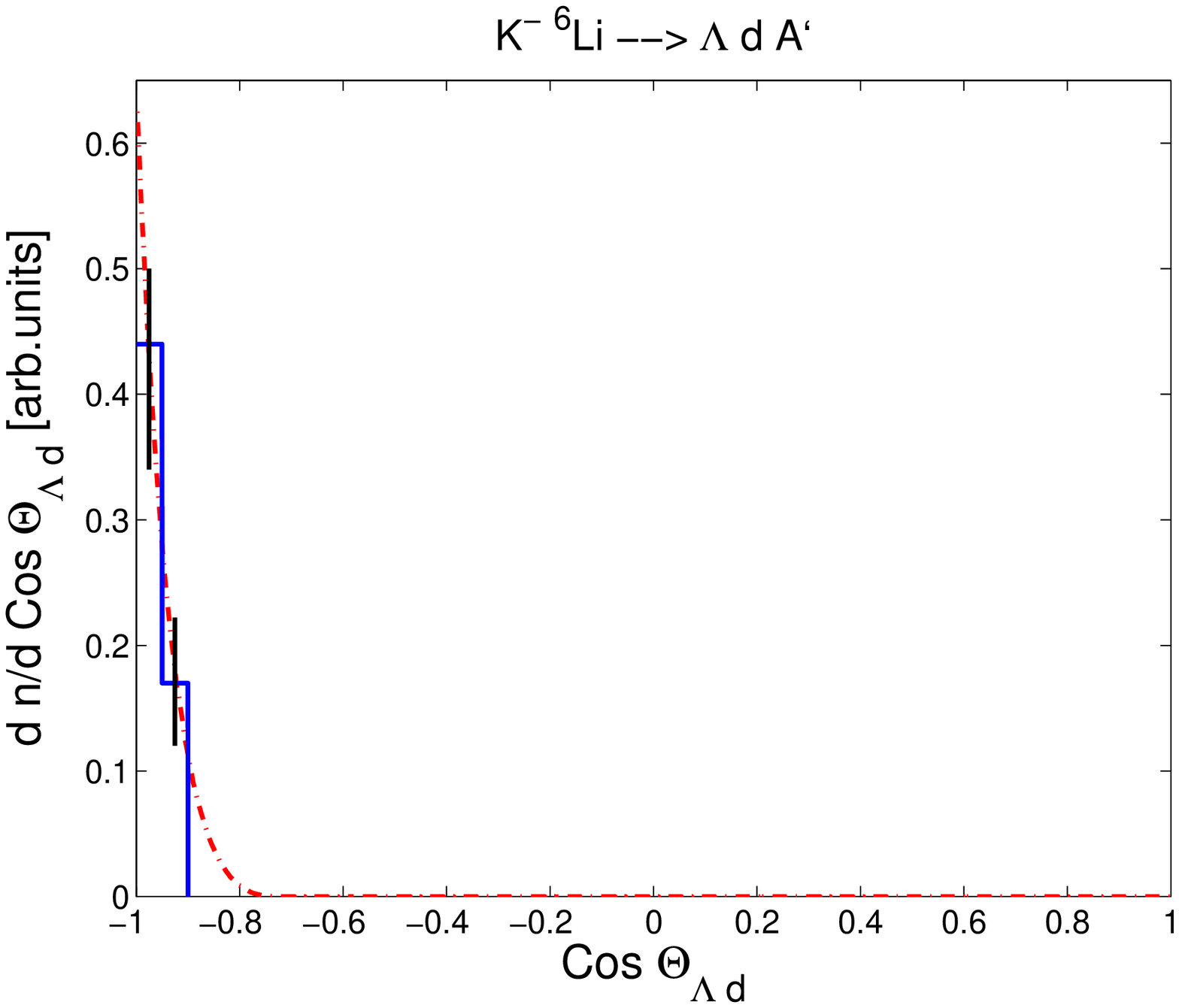} 
\caption{The $\Lambda d$ invariant mass distribution (left plot) and 
$\Lambda d$ angular distribution (right plot) for $K^-$ absorption in 
$^{6}$Li. Histogram and error bars are from the experimental paper 
\cite{:2007ph}, while the dot-dashed curve is the result of our 
calculation \cite{Magas:2008bp}.} 
\label{fig2} 
\end{figure*}

In Ref. \cite{Magas:2008bp} we performed detailed calculations of 
$K^-$ absorption by three nucleons in $^6$Li and showed that all features 
observed in \cite{:2007ph} could be well interpreted in the picture of three 
body kaon absorption, as suggested in \cite{Suzuki:2007kn}, 
with the rest of the nucleons acting as spectators - see Fig. \ref{fig2}.  
It is also important to note that $ ^{12}$C was also used as a target 
in the same FINUDA experiment, and the corresponding $\Lambda d$ invariant 
mass spectrum does not show a clear peak \cite{:2007ph}. This was attributed 
in \cite{:2007ph} to FSI of the particles produced in the primary absorption 
process with the residual nucleus, in complete agreement with the mechanisms 
discussed in Ref. \cite{Magas:2006fn}.

\section{Conclusions} 
\label{concl}

We have shown that the peaks observed by FINUDA in the 
($\Lambda p$) \cite{Agnello:2005qj} and ($\Lambda d$) \cite{:2007ph} 
invariant mass distributions, following the absorption of stopped $K^-$ 
in different nuclei, are naturally explained in terms of: \\ 
\noindent
- $K^-$ absorption by two \cite{Magas:2006fn} or three 
\cite{Magas:2008bp} nucleons correspondingly, leaving the rest of the 
target nucleus as spectator; \\
- for the reactions on heavy nuclei, the subsequent interactions of the 
particles produced in the primary absorption process ($\Lambda$, $p$, $d$, 
etc.) with the residual nucleus have to be taken into account. \\
Thus, at present, there is no experimental evidence of deeply bound $K^-$ 
states in nuclei. 

\section{Acknowledgements}
The authors thank H. Toki for fruitful and enlightening discussions.
This work is partly supported by contracts FIS2006-03438 and
FIS2005-03142 from MEC (Spain) and FEDER, the Generalitat de
Catalunya contract 2005SGR-00343 and by the Generalitat Valenciana.
We also acknowledge the support of the European Community-Research
Infrastructure Integrating Activity "Study of Strongly Interacting Matter" 
(HadronPhysics2, Grant Agreement n. 227431) under the EU 7th Framework 
Programme.


\begin{thebibliography}{0} 

\bibitem{Ramos:2008npa} A.~Ramos, V.~K.~Magas, E.~Oset and H.~Toki, 
Nucl.\ Phys.\ {\bf A804}, 219 (2008). 

\bibitem{Agnello:2005qj} M.~Agnello {\it et al.}  [FINUDA Collaboration], 
Phys.\ Rev.\ Lett. {\bf 94}, 212303 (2005).

\bibitem{Magas:2006fn} V.~K.~Magas, E.~Oset, A.~Ramos and H.~Toki, 
Phys.\ Rev.\ {\bf C74}, 025206 (2006). 

\bibitem{future_finuda} T. Bressani, private communication.

\bibitem{:2007ph} M.~Agnello {\it et al.}  [FINUDA Collaboration], 
Phys.\ Lett.\ {\bf B654}, 80 (2007). 

\bibitem{Suzuki:2007kn} T.~Suzuki {\it et al.} [KEK-PS E549 Collaboration], 
Phys.\ Rev.\ {\bf C76}, 068202 (2007). 

\bibitem{Magas:2008bp} V.~K.~Magas, E.~Oset and A.~Ramos,
Phys.\ Rev.\ {\bf C77}, 065210 (2008). 

\end{thebibliography}
\end{document}